\documentstyle[prl,aps,latexsym,epsfig]{revtex}

\draft
\newcommand{\be}{\begin{eqnarray}}
\newcommand{\ee}{\end{eqnarray}}
\newcommand{\QM }{quantum mechanics}

\title{Violations of local realism by two entangled qu$N$its are 
stronger than for two qubits}
\author{Dagomir Kaszlikowski$^1$, Piotr Gnaci\'nski$^1$, Marek \.Zukowski
$^{1,2}$, Wieslaw Miklaszewski$^1$ and Anton Zeilinger$^2$ }
\address{$^1$Instytut Fizyki Teoretycznej 
i Astrofizyki Uniwersytet
Gda\'nski, PL-80-952 Gda\'nsk, Poland,
$^2$Institut f\"ur Experimentalphysik, Univesit\"at Wien, Austria} 
\begin{document}
\maketitle
\begin{abstract}
Tests of local realism vs quantum mechanics based on Bell's inequality
employ two entangled qubits. We investigate the  general case of two entangled
qu$N$its, i.e. quantum systems defined in an $N$-dimensional Hilbert 
space.
Via a numerical linear optimization method we show 
that
violations of local realism are stronger for two maximally entangled 
qu$N$its ($3\leq N\leq 9$), than for two qubits and that they increase with $N$.
The two qu$N$it measurements can be experimentally 
realized using entangled photons and unbiased multiport 
beamsplitters.
\end{abstract}
\twocolumn

John Bell \cite{bell}  has shown
that no local realistic models 
can agree with all quantum mechanical predictions
for the maximally
entangled states of two two-state systems (qubits).
After some years researchers started to ask questions about the
Bell theorem for more complicated systems. The most surprising
answer came from the  GHZ theorem \cite{GHZ}:
 for three or more qubits
 the conflict
between local realism and quantum mechanics 
is much sharper than for two qubits. 
The other possible extension are entangled states
of pairs of $N$-state systems, qu$N$its, with $N\geq 3$. First results, 
in 1980-82, suggested that
the conflict between local realism and quantum mechanics
diminishes with growing N \cite{MERMIN}. 
This was felt to be in concurrence with the old
quantum wisdom of higher quantum numbers leading to a
quasi-classical behavior.
However, that  early research was confined to Stern-Gerlach type
measurements performed on pairs of entangled ${N-1\over2}$
spins \cite{MERMIN}. Since operation of a Stern-Gerlach device
depends solely on the orientation of the quantization axis,
i.e. on only two parameters, devices of this kind
cannot make projections into arbitrary states of the subsystems. 
That is, they cannot make full use of 
the richness of the $N$-dimensional Hilbert space.

In early 1990's Peres and Gisin \cite{PERES} 
considered certain {\em dichotomic} observables applied to maximally
entangled pairs of qu$N$its. They showed that
 the violation of local realism, or
more precisely of the CHSH inequalities, survives
the limit of $N\rightarrow\infty$, but never exceeds the violation 
by two qubits, in agreement with Cirel'son limit
 \cite{TSIRELSON}, i.e. it is limited by the 
factor of $\sqrt{2}$.  
Therefore, the question 
whether the violation of local realism increases or not
  with growing $N$ for general observables 
was still left open.

To answer this question it is necessary first to adopt  an objective measure
of the magnitude of violation of local realism.
To this end, consider  two qu$N$it systems described by mixed states 
in the form of 
\begin{eqnarray}
&\rho_N(F_{N})=F_{N}\rho_{noise}+
(1-F_{N})|\Psi_{max}^N\rangle\langle\Psi_{max}^N|,&
\label{werner}
\end{eqnarray}
where the positive parameter $F_N\leq 1$ determines the "noise
fraction" within the full state, $\rho_{noise}=\frac{1}{N^2}\hat{I}$,
and $|\Psi_{max}^N\rangle$ is a maximally entangled two
qu$N$it state, say
\begin{eqnarray}
&&|\Psi_{max}^N\rangle={1\over\sqrt N}\sum_{m=1}^{N}|m\rangle_A|m\rangle_B.
\label{state}
\end{eqnarray}
In (\ref{state}) $|m\rangle_A$ ($|m\rangle_B$) describes particle 
$A$ ($B$) in its mode $m$. One has 
${}_x\langle m| m'\rangle_x=\delta_{m,m'}$, with $x=A,B$.
The threshold maximal $F_{N}^{max}$,
for which the state $\rho_N(F_{N})$ still does not  allow a local realistic model,
will be our  value of the strength of violation of local realism.
The higher $F_N^{max}$ the higher noise admixture will be 
required 
to hide the non-classicality of the quantum prediction.
In experiments  the visibility 
parameter $V$, effectively
equivalent to
$1-F_{N}$,  
is 
the 
usual measure of the reduction of interferometric contrast (visibility).

We shall study the case of
two observers Alice and Bob performing measurements of 
local non-degenerate observables, each on her/his qu$N$it of an
entangled pair in the state $\rho_{N}(F_N)$. 
Let us imagine that Alice can 
choose between 
two non-degenerate observables $A_1$ and $A_2$, and that each  observable is 
defined such 
that it has the full spectrum characterized by
all integers from $k=1$ to $N$.
Bob can choose between
$B_1$ and $B_2$, both with the same spectrum as above ($l=1,2,...,N$).
Thus, the observers can perform $2\times 2$ 
mutually exclusive global experiments. 
The quantum probability  for the specific pair of results, $k$ for
Alice
and $l$ for Bob, 
provided a specific pair of local observables is
chosen, $A_i$ by Alice and $B_j$ by Bob, will be denoted by $
P^{QM}_{F_N}(k;l|{A}_i,{B}_j)$. 
Quantum mechanics makes predictions for  the complete set of $4N^2$ 
such probabilities, and nothing more.

The hypothesis of  local hidden variables tries to go beyond.
The basic assumption there is that each particle carries a 
probabilistic or deterministic set of instructions how to respond to all 
possible local measurements it might be subject to. Therefore local realism
assumes the existence of  non-negative joint
probabilities  involving all 
possible observations from which it should be possible to 
obtain all the quantum predictions as marginals (see, e.g. \cite{FINE}, \cite{PERESBELL}).
Let us denote these hypothetical probabilities by
$P^{HV}(k,m;l,n|A_1,A_2,B_1,B_2)$, where $k$ and $m$,
represent the outcome values for Alice's observables ($l$ and $n$
for Bob's). 
In \QM\ one cannot even define such objects, since 
they involve mutually incompatible measurements.
The local hidden variable probabilities for the experimentally observed events,
$k$ ($m$) by Alice measuring $A_1$ ($A_2$), and $l$ ($n$) by Bob 
measuring $B_1$ ($B_2$), are 
the marginals
\be
&P^{HV}(k;l|A_1, B_1)=
\sum_{m}
\sum_{n}
P^{HV}(k,m;l,n),&\nonumber \\
&P^{HV}(k;n|A_1,B_2)=
\sum_{m}
\sum_{l}
P^{HV}(k,m;l,n),&\nonumber \\
&P^{HV}(m;l|A_2,B_1)=
\sum_{k}
\sum_{n}
P^{HV}(k,m;l,n),&\nonumber \\
&P^{HV}(m;n|A_2,B_2)=
\sum_{k}
\sum_{l}
P^{HV}(k,m;l,n),&\nonumber \\ 
\label{sumation}
\ee
where $P^{HV}(k,m;l,n)$ is a short hand notation for
$P^{HV}(k,m;l,n|A_1,A_2,B_1,B_2)$. 
The $4N^2$ 
equations (\ref{sumation})
form the full set of necessary and sufficient conditions
for the existence of  
local realistic
description of the experiment, i.e., for the joint
probability distribution $P^{HV}(k,m;l,n)$.
The Bell Theorem says that 
certain predictions by quantum mechanics
are in conflict  with the local hidden variable model (\ref{sumation}).
Evidently, the conflict disappears when enough noise is added, as in the state
(\ref{werner}), since that noise has a local realistic model. Therefore 
a threshold $F_N^{max}$ exists
below  which one cannot have any local realistic model with 
$P^{HV}(k;l|A_i,B_j)=P^{QM}_{F_N}(k;l|A_i,B_j)$.
Our goal is to find observables for the two qu$N$its 
returning the highest possible critical $F_N^{max}$.

Up to date, no one has derived Bell-type inequalities
that are necessary and sufficient 
conditions for (\ref{sumation}) to hold, with the exception of 
the $N=2$ case  (see \cite{PERESBELL}).
However there are numerical tools, in the form of the very 
well developed theory and methods of linear optimization, which are
perfectly suited for tackling exactly such problems \cite{ZKBL}.

The quantum probabilities, when the state is given by (\ref{werner}), have 
the following structure
\begin{eqnarray}
&P^{QM}_{F_N}(k;l|A_i,B_j)& \nonumber \\
&=\frac{1}{N^2}F_N+(1-F_N)P^{QM}(k;l|A_i,B_j),& \\
\nonumber
\end{eqnarray} 
where $P^{QM}(k;l|A_i,B_j)$ is the probability for the given pair 
of events for the pure maximally entangled state.
The set of conditions (\ref{sumation}) with $P^{QM}_{F_N}(k;l|A_i,B_j)$
replacing $P^{HV}(k;l|A_i,B_j)$ imposes linear constraints on the $N^{4}$
``hidden probabilities" $P^{HV}(k,m;l,n)$ 
and on the parameter $F_N$, which are the nonnegative unknowns. We have more  
unknowns ($N^{4}+1$) than equations 
($4 N^2+1$, with the normalization condition for the hidden probabilities), and we want to find the
minimal $F_N$ for which the set of constraints can still be satisfied.
This is a typical linear optimization problem for which
lots of excellent algorithms exist.  We have used the
state-of-the-art algorithm HOPDM 2.30.
(Higher Order Primal Dual Method) \cite{GONDZIO1}.
 It is important to stress that 
for cross-checking  four independently written codes were used,
one of them employing a different linear optimization procedure
(from the NAG Library). 

We were interested in finding
such observables for which the threshold $F_N$ acquires
the highest possible value. 
To find 
optimal sets  of observables we have used a numerical procedure based
on the downhill simplex method (so called amoeba)
\cite{Recipies}. If the 
dimension of the domain of a function is $D$ (in our case $D=4n$,
where $n$ is the number of parameters specifying the nondegenerate 
local observables
belonging to a chosen family),
the procedure first randomly generates
$D+1$ points. In this way it creates the vertices of a
starting simplex. Next it calculates the value of the
function at the vertices and
starts exploring the space by stretching and contracting the simplex. 
In every
step, 
when it 
finds vertices where the value of the function is higher than in others, it 
"goes" 
in this direction (see e.g.
\cite{Recipies}).

Let us now move to the question of finding a family of observables,
which returns critical $F_N$'s that are above the well known threshold
for the two qubit case, $1-\frac{1}{\sqrt{2}}$. As it was said earlier,
and was confirmed by our numerical results, 
Stern-Gerlach type measurements are not suitable. More exotic 
observables are needed.

First we discuss how experiments
on two entangled qu$N$its might be performed. In view of 
the unavailability of higher spin entanglement it is fortunate
that qu$N$it entanglement can be studied exploiting 
momentum conservation in the many processes of two-particle generation,
most notably in the parametric down conversion generation of entangled
photon pairs. This results in strong correlations between the propagation directions of the particles in a pair. One can then submit $N$ spatial modes
of each particle to a multiport beamsplitter \cite{MULT}.

Application of multiports in the context of quantum
entanglement has been first  discussed by Klyshko \cite{KLYSHKO}.
Proposals of Bell experiments with 
the multiports were presented in \cite{ZEILINGER93},
and further developed  in  \cite{MULT}.
Multiport devices can
reproduce all finite dimensional unitary transformations for
single-photon states \cite{RECK94}, therefore they are characterized by $N^2-1$
real parameters. 

In order to limit computer time we restricted our
analysis to unbiased multiports \cite{MULT}, more specifically to Bell multiports.
Unbiased multiports have the property that a photon entering into
any {\it single} input port (out of the $N$), has equal chances to exit from any output port.
In addition, for Bell multiports \cite{MULT}  
the elements of their  unitary transition matrix, ${\bf U}^{N}$, are 
{\it solely} powers of the N-th root of unity
$
\gamma_N=exp{(i2\pi/N)},
$
namely
$
{\bf U}^{N}_{ji}= \frac{1}{\sqrt{N}}\gamma_N^{(j-1)(i-1)}
$.

Let us now imagine two spatially separated experimenters who
perform the experiment of FIG. 1. (described in the caption). The 
initial maximally entangled state (\ref{state}) of the two qu$N$its
can be prepared with the aid of parametric down conversion (see
\cite{MULT}).
The two
sets of phase shifters at the inputs of the multiports 
(one phase shifter in
each beam) introduce phase factor $e^{i(\phi^m_A+\phi^m_B)}$ in
front of the $m$-th component of the state
(\ref{state}), where $\phi^m_A$ and $\phi^m_B$
denote the local phase shifts.
\begin{figure}[htbp]
\begin{center}
\includegraphics[angle=0, width=8.5cm]{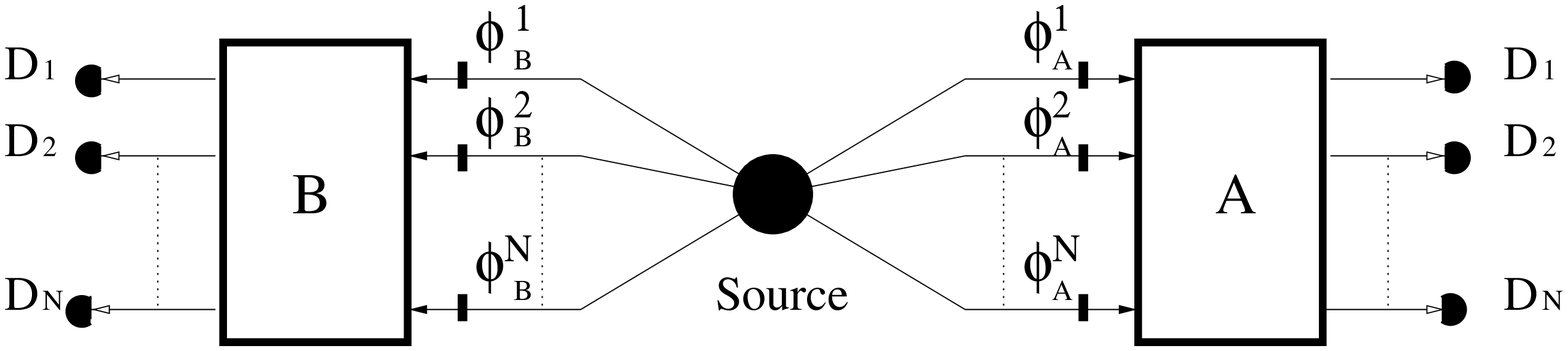}
\caption{The experiment of Alice and Bob with entangled qu$N$its. 
Each of their measuring apparata consist of a set of $N$
phase shifters just in front of an $2N$ port Bell multiport, and
$N$ photon detectors $D_k, D_l$ (perfect, in the gedanken situation described
here) which register photons in the output ports of the device. 
The phase shifters serve the role of the devices which
set the free macroscopic, classical parameters that can be
controlled by the experimenters. The source 
produces a beam-entangled two particle
state.}  
\label{plot1}
\end{center} 
\end{figure}

Each set of local phase shifts constitutes the
interferometric realizations of the "knobs" at the disposal of
the observer controlling the local measuring apparatus, which
incorporates also the Bell multiport and N detectors. 
In this way the local observable is defined. Its eigenvalues refer simply 
to registration at one of the $N$ detectors behind the multiport. The
quantum prediction for the joint probability $P^{QM}_{F_N}(k,l)$ to
detect a photon at the $k$-th output of the multiport A and
another one at the $l$-th output of the multiport B is given by
\cite{MULT}:
\begin{eqnarray}
&P^{QM}_{F_N}{(k,l;\phi^1_A,...\phi^N_A,\phi^1_B,...\phi^N_B )}= 
\frac{F_N}{N^2}& \nonumber \\
&+\frac{1-F_N}{N}\left|\sum^N_{m=1}\exp{[i(\phi^m_A+\phi^m_B)]}
{\bf U}^N_{mk}{\bf U}^N_{ml}\right|^2&
\nonumber \\
& =(\frac{1}{N^3})\left(N+
2(1-F_N)\sum^N_{m>n}\cos{({\bf \Phi}^m_{kl}-{\bf \Phi}^n_{kl})}\right),& 
\label{25a}
\end{eqnarray}
where
${\bf \Phi}^m_{kl} \equiv \phi^m_A+\phi^m_B +[m(k+l-2)] \frac{2\pi}{N}$.
The counts at a single detector, of course are constant, and 
do not depend upon
the local phase settings:
$
P^{QM}_{F_N}{(k)}=P^{QM}_{F_N}{(l)}={1}/{N}.$

The numerical values of the  threshold
$F_N$ are given in fig. 2.
\begin{figure}[htbp]
\begin{center}
\includegraphics[angle=270, width=8.5cm]{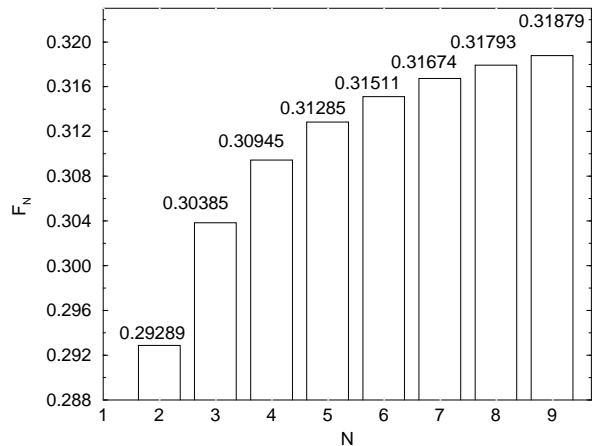}
\caption{Maximal fraction $F^{max}_N$ of pure noise admixture to a
maximally entangled two qu$N$it system, such that a local 
realistic explanation still cannot be upheld. For smaller noise fractions
a conflict arises between quantum mechanics and local realism.
The result for $N=2$ agrees with the standard threshold of $1-{1\over\sqrt 2}$.}  
\label{plot2}
\end{center} 
\end{figure}
It is evident, that indeed two entangled
qu$N$its violate local realism stronger than two entangled qubits, 
and that the violation increases monotonically with $N$.
It is tempting to contemplate the limit of  $N\rightarrow\infty$.
While obviously the values of $F_N^{max}$ seem to saturate, at present 
we cannot give a definite asymptotic value. 

A few words of comment are needed.  
One may argue that because of the rather large number
of local macroscopic parameters (the phases)
defining the function to be maximized with the amoeba we 
could have missed the global minimum. 
While this argument cannot be ruled out in principle, we 
stress that in that case the ultimate violation would even be larger.
This would only strengthen our conclusion that two entangled qu$N$its
are in stronger conflict with local realism than two entangled
qubits.

Based on the numerical results, i.e. the values of the optimal phase settings,
and on the structure of the local hidden variable model for $F_3^{max}$,
an algebraic calculation was performed \cite{KASZLIKOWSKI}  
showing that for the two qutrits $(N=3)$ 
experiment the exact value for $F_3^{max}$ is $\frac{11-6\sqrt{3}}{2}$.
One should also mention that 
for two spin 1 particles in a singlet state observed by
two Stern Gerlach apparatuses our method gives 
$F^{SG}_3=0.1945$, which is much 
smaller than $1-\frac{1}{\sqrt{2}}$, confirming that 
such measurements are not optimal in the sense of leading to 
maximal possible violations of local realism.

An important question is whether  unbiased Bell  multiports provide us with 
a family of observables in maximal conflict with 
local realism. For a check of this question
we have also calculated the threshold value
of $F_3$ for the case where both observers apply to
the incoming qutrit the most general
unitary transformation belonging to a full SU(3) group (i.e. 
we have {\it any} trichotomic observables on each side). Again
we have assumed that each observer chooses between two sets of
local settings. However, in this case each set consists of 8
local settings rather than the three (effectively two) in the tritter case.
The result appears to be the same as for two tritters.
While this might suggest 
that for $N=3$ Bell
multiports are optimal devices to test \QM\ against
local realism, this needs to be further investigated.

It is interesting to compare our results
with the limit 
for the non-separability of the density matrices (\ref{werner}).
 The critical minimal 
$F_{N}$ 
for which a density matrix 
(\ref{werner}) is separable is ${N\over N+1}$ (see  \cite{HOROD}).  
The fact that this limit is always higher than 
ours indicates that the  requirement of having a local 
quantum description of the two subsystems is a much more stringent
condition than the requirement of admitting any possible 
local realistic model.

It will be interesting to consider within our approach 
different families of states, generalizations to more than two particles, 
extensions of the families of observables and 
to see if more than two (e.g. $A_1,A_2,A_3$)
experiments performed on either
side can lead to even stronger violations
of local realism.
The questions concerning the critical $F_{N}$ are also important in
the attempts to generalize Ekert's quantum cryptographic protocol
to qutrits and higher systems \cite{TRITCRYPTO}. 

Our method is numerical,
and is based on linear optimization.
It is a development of the approach of \cite{ZKBL}. 
The exploding (with $N$)
difficulty of approaching this type of problems via algebraic-analytical 
methods (generalized Bell inequalities, via the Farkas lemma, etc.) 
has been  exposed 
in \cite{PERESBELL}.

It will certainly be fascinating to see laboratory realizations of
the experimental schemes discussed here.

We thank Jacek Gondzio (Edinburgh) for courtesy in 
allowing to use his most recent version of the code HOPDM. We
also thank Adam Baturo and Jan-\AA ke Larsson for their
contribution to the two qubit stage of the project \cite{ZKBL}.
The work is supported by the Austrian-Polish program 24/2000
Quantum Communication and Quantum Information. Additional support:
AZ was supported by the Austrian FWF project F1506;
MZ was supported by the University of Gdansk Grant No
BW/5400-5-0032-0 and The Erwin Schr\"odinger
International Institute for Mathematical
Physics, Vienna;  DK was supported by Fundacja na Rzecz Nauki
Polskiej and
the KBN Grant 2 P03B 096 15.

The paper is dedicated to the memory of the late 
Professor David N. Klyshko,
our Friend (Z \& \.Z) and a great innovator in the field of nonlinear 
quantum optics.

\end{document}